\documentstyle[preprint,aps,prb]{revtex}
\begin{document}
\draft

\title{Spin-Peierls and Antiferromagnetic Phases in Cu$_{1-x}$Zn$_x$GeO$_3$:\\
A Neutron Scattering Study}

\author{Michael C. Martin\thanks{Present address:  Department of Physics, U.C.
Berkeley, Berkeley, CA  94720-7300},  M. Hase\thanks{Present address: 
National Research Institute for Metals,
1-2-1 Sengen, Tsukuba, Ibaragi 305, Japan}, K. Hirota\thanks{Present address:   
Department of Physics, Tohoku University, Sendai 980-77, Japan},  and G. Shirane}
\address{Department of Physics, Brookhaven National Laboratory, Upton, NY 
11973-5000}
\author{Y. Sasago, N. Koide and K. Uchinokura}
\address{Department of Applied Physics, University of Tokyo, Bunkyo-ku,  Tokyo
113, Japan}

\date{\today}
\maketitle

\begin{abstract}
Comprehensive neutron scattering studies were carried out on a series of 
high-quality single crystals of 
Cu$_{1-x}$Zn$_x$GeO$_3$.  The Zn concentration, $x$, was determined for 
each sample using Electron Probe Micro-Analysis.  The measured Zn 
concentrations were found to be 40-80\% lower than the nominal values.  
Nevertheless the measured concentrations cover a wide range which enables
a systematic study of the effects due to Zn-doping.   We have confirmed
the coexistence of spin-Peierls (SP) and antiferromagnetic (AF)
orderings at low temperatures and the measured phase diagram is presented.  
Most surprisingly, long-range AF ordering occurs even in the lowest available 
Zn concentration, x=0.42\%, which places important constraints on theoretical 
models of the AF-SP coexistence.  Magnetic excitations are also examined in
detail.  The AF excitations are sharp at low energies and show no 
considerable broadening as $x$ increases indicating that the AF ordering 
remains long ranged for $x$ up to 4.7\%.  On the other hand, the SP phase
exhibits increasing disorder as $x$ increases, as shown from the broadening of
the SP excitations as well as the dimer reflection peaks.
\end{abstract}

\section{Introduction}

In 1993 the discovery of an inorganic spin-Peierls (SP) compound, CuGeO$_3$,
was first reported by Hase {\it et al.} \cite{hase93}  This material has
chains of Cu$^{2+}$ ($S=1/2$) along its $c$-axis which
distort into dimers below the SP transition temperature,
$T_{SP}=14.2$K.\cite{hirota94}   
The dimerized structure has a singlet ground state
and a triplet excited state; singlet-to-triplet excitations were
demonstrated by 
Nishi {\it et al.}\cite{nishi94} using inelastic neutron scattering.
CuGeO$_3$ is interesting not only because it is the first known inorganic
spin-Peierls system, but also because evidence is mounting that it does not
behave like a typical SP system.  It has been believed that good 
one-dimensionality was needed for a material to undergo a spin-Peierls
transition.  However nearest-neighbor exchange parameters, estimated
from the dispersions of magnetic excitations\cite{nishi94}
imply that the intrachain coupling is only 10 times stronger than the
interchain coupling, which is far smaller than is found for other 
one-dimensional systems, typically 10$^2$ to 10$^4$
[\onlinecite{buyers},\onlinecite{NENP}].  Recent neutron-scattering 
measurements have revealed that the softening of a zone-boundary phonon as 
expected in ordinary SP systems does not exist.\cite{hirota95}  
In addition, high-pressure neutron scattering
studies\cite{katano,nishipress} have shown that as the pressure increases,
$T_{SP}$ and the SP energy gap increase while the lattice dimerization 
decreases, implying an additional mechanism, such as a spin-only effect, 
is at work.
The susceptibility above $T_{SP}$\cite{hase93} and the critical exponent 
$\beta$ of the SP energy gap\cite{martin,lussier96} both deviate 
substantially from the theoretical calculations of Bonner and 
Fisher\cite{bonner} and Cross and Fisher\cite{cross} respectively, which
work well for organic SP systems.
Castilla {\it et al.}\cite{castilla}  have recently
proposed one possible theoretical model 
where the magnetic properties of CuGeO$_{3}$ can be well described by the
one-dimensional Heisenberg model with competition between nearest-neighbor and
next-nearest neighbor antiferromagnetic interactions.

Shortly after the discovery\cite{hase93} of CuGeO$_3$, a series of extensive
studies were begun on samples with small amounts of dopant.  Investigations 
have concentrated on systems where Cu atoms were replaced with Zn 
\cite{hase93-2,hase95,oseroff,lussier,hase96,sasago} or Ge was replaced with 
Si.\cite{renard,poirier,regnault}   It is now well established that a new
antiferromagnetic (AF) ordered phase appears at low temperatures in both 
Zn-doped\cite{hase93-2,sasago} and Si-doped\cite{regnault} samples.  A
preliminary phase diagram was determined from susceptibility 
measurements on powder samples\cite{hase93-2} which showed that $T_{SP}$
decreases with increasing Zn concentration while the N\'eel temperature
increases to $T_N \sim$ 4K.  This phase diagram was recently confirmed 
using single crystals\cite{hase95,sasago} and in Figure \ref{phase_diagram}
we present the phase diagram as determined from neutron and susceptibility
measurements on all single crystals produced for the present study.
The most intriguing feature reported so far in these doped CuGeO$_3$ systems
is the coexistence of the SP and AF states below the N\'eel temperature. 
This was first demonstrated by 
Regnault {\it et al.}\cite{regnault} who used neutron scattering to 
observe the SP order parameter, i.e. the lattice dimerization, become smaller
but still finite below $T_N$.  
Sasago {\it et al.}\cite{sasago} showed  that the SP and AF states 
also coexist for samples having a nominal range of 2 -- 6\% Zn.
The coexistence of a SP singlet state and AF ordering
was quite a surprise since it was thought that these two ordered states were
mutually exclusive, and it adds another puzzle to the already unusual SP
state in CuGeO$_3$.  Recently, Fukuyama {\it et al.}\cite{fukuyama} have
proposed a theoretical model for disorder-induced antiferromagnetic long
range order within  a spin-Peierls system.  This model concludes that an
infinitesimal amount of impurity can induce a long-range AF state in 
a spin-gap system (which includes spin-ladders and two dimensional systems,
as well as SP systems).   The comparison of this theory with our experimental 
results will be presented.

In the present paper comprehensive neutron scattering results are presented 
studying Zn-doped CuGeO$_3$ single crystals grown by the floating-zone (FZ)
method.  The amount of Zn-dopant, $x$, was carefully determined using 
Electron Probe Micro-Analysis (EPMA) for each crystal.  The SP dimerization
and the AF magnetic moment are surveyed as a function of $x$.  Magnetic 
excitations in both the SP and the SP-AF coexistence states were studied
in detail as a function of temperature, $q$, and $x$ yielding some important
conclusions about the interplay of the AF and SP states.  

\section{Experimental Details}

\subsection{Sample Preparation} 
A series of relatively large ($\sim 0.4$~cm$^3$)
Cu$_{1-x}$Zn$_x$GeO$_3$ single crystals having {\em nominal} Zn concentrations
of 1, 2, 4, 6, and 8\% were grown using the floating-zone (FZ) method.  
The crystals grow along the $c$-axis (the chain direction) and typically 
had mosaic spreads of less than 0.3 degrees.
As shown in Figure \ref{samples}, the crystals were cut into several
parts for characterization using Electron Probe Micro-Analysis (EPMA) 
and measurements of neutron scattering and susceptibility.
Typical susceptibilities along the $c$ direction for several values of $x$ 
are shown in the lower part of Figure \ref{samples}.
The EPMA analysis revealed that the true Zn concentrations were consistently 
lower than the nominal values:
the average measured values of $x$ are 0.42, 0.9, 3.2, 2.1, and 4.7\% 
corresponding to the nominal values of 1, 2, 4, 6, and 8\%, respectively.

As the Zn concentration increases,
it becomes more difficult to grow a single crystal with a single domain.
In order to obtain high quality single crystals with small mosaic spreads,
the speed at which the zone is moved must be kept slow.  On the other 
hand, impurities (Zn atoms) tend to be purged from the melted zone if the 
growth process is too slow.
It is thus difficult to fabricate crystals both of high quality and with 
high Zn concentration.  
We were able to grow crystals with a fairly wide range of Zn concentrations
enabling a systematic study of the effects of Zn doping.  
The homogeneity of the Zn concentrations was checked by measuring various
regions ($2\times 2\ \mu$m$^2$ each) of the samples by EPMA.  These
homogeneities are quoted as uncertainties in
Table \ref{tab} which also presents the crystal volumes and measured
transition temperatures.  Both SP and AF transitions are quite sharp, 
consistent with small error bars in the actual Zn concentrations.

\subsection{Neutron Scattering Spectrometers}

Neutron scattering measurements were carried out using triple-axis
spectrometers on the H7 and H8 (14.7 meV
incident energy neutrons), and H9a (5 meV) beamlines of the 
High Flux Beam Reactor at Brookhaven National Laboratory.  The crystals 
were mounted in helium filled aluminum cans which were subsequently  
attached to the cold finger of a cryostat.  The samples were aligned 
so as to  place the $(0\ k\ l)$ or $(h\ k\ h$) zones in the experimental
scattering  plane.  Incident neutrons with energies of 14.7 or 5.0~meV were
selected by  a pyrolytic graphite (PG) $(0\ 0\ 2)$ monochromator;
PG $(0\ 0\ 2)$ was  also used for an analyzer.   
To eliminate higher-order harmonics, PG
filters were placed before the sample  and after the analyzer.  The beam was
horizontally collimated, typically  40$'$-40$'$-Sample-40$'$-80$'$ in sequence
from the reactor core to the detector. Tighter collimations were used for some
of the higher resolution measurements presented.
When comparing the intensities of dimerization
and magnetic superlattice peaks between the various crystals, Bragg peaks
unaffected by extinction are required for use as normalization.  
The weak $(0\ 2\ 1)$ and $(1\ 2\ 1)$ peaks fill this need and are used
for our data analysis.

\section{Order Parameters}
\subsection{Phase Diagram}

The transition temperatures $T_{SP}$ and $T_N$, where the sample enters 
the SP and AF states respectively, 
were determined for each crystal using both magnetic susceptibility 
measurements and observing critical scattering peaks in the neutron studies.
Figure \ref{order_parameters} plots the intensity of the SP and AF 
superlattice peaks as a function of temperature for 3.2 and 0.42\%
Zn-doped samples.
$T_{SP}$ and $T_{N}$ for each crystal are summarized
in Table \ref{tab}.  The full phase diagram including susceptibility data
from numerous crystals is presented in Figure \ref{phase_diagram}.  

The most significant result here is that AF ordering is observed for even the 
smallest available amounts of Zn, $x=0.42\%$.  As shown in Figure
\ref{order_parameters}(b), the AF superlattice peak is nearly resolution
limited demonstrating that a long-ranged N\'eel ordering is induced by
very small amounts of Zn impurities.  A second interesting result is that 
$T_{SP}$ and $T_N$ change as $x$ is increased to about 2\%, but then do not
change much for Zn concentrations exceeding 2\%.

\subsection{Dimerization}

Hirota {\it et al.}\cite{hirota94} determined that as the dimerization 
occurs below $T_{SP}$, 
the $({h\over{2}}\ k\ {l\over{2}})$ superlattice peaks, with
$h$ and $l$ odd and $k$ any integer, appear. 
The intensities of the SP superlattice peaks decrease continuously as a
function of increasing Zn content indicating a weakening SP dimerization.   
The transition is less sharp in the Zn-doped crystals compared to pure  
CuGeO$_3$, and it can be estimated by fitting the data with 
a power law model which includes a Gaussian distribution of transition 
temperatures.  

Using the maximum intensity of these SP superlattice peaks (at about 6~K), 
the dimerizing atomic displacement, $\delta$, can be calculated as compared to
the displacement in pure CuGeO$_3$, which we define as $\delta_0$.  To 
normalize with respect to crystal volume, 
the extinction-free (1 2 1) Bragg peak is used as a reference for
each sample.  The structural factor $F$ is calculated for each peak using the 
relation $F=\sqrt{I\sin 2\theta}$, where $I$ is the measured intensity and 
$\theta$ is the scattering angle.  The approximate atomic displacements 
are now calculated for each dopant level  by comparing the ratio
$F_{\rm obs}({1\over{2}}\ 3\  {1\over{2}}) / F_{\rm obs}(1\ 2\ 1)$ for each
Zn-doped sample to the known ratio in the undoped case.  The numerical 
results are 
presented in Table \ref{tab}.  As the Zn-doping increases, the average
atomic distortion is reduced dramatically; by adding only 1\% Zn the 
dimerizing atomic shift loses 1/3 of its magnitude. 

\subsection{Antiferromagnetic moment}

Hase {\it et al.}\cite{hase93-2} found that the AF ordering is associated
with a superlattice
reflection at $(0\ 1\ \frac{1}{2})$.
The magnetic moment of the AF state can be calculated from the observed
intensity of this peak.  Using the (0 2 1) peak as an
extinction-free normalization, we compare the 
$F_{\rm obs}(0\ 1\ {1\over{2}}) / F_{\rm obs}(0\ 2\ 1)$ ratio to 
the ratio previously measured by Hase {\it et al.}\cite{hase96} 
for a different 3.4\% Zn-doped crystal.  Hase {\it et al.} calculated t
he magnetic moment, $\mu_{eff}$, extrapolated 
to zero temperature for that sample to be 
$\mu_{eff} = 0.22\mu_{B}$.  By comparing the normalized superlattice
peak intensities, values for the magnetic moments of each of the
Zn-doped samples are obtained, and are summarized in 
Table \ref{tab}.  In all Zn-doped samples studied the 
SP dimer superlattice peak loses some intensity below $T_N$.  Since the
superlattice peak intensity is directly  related to the atomic displacements 
($I \propto \delta^2$) this observation  shows that the dimers are indeed 
adversely affected by the onset of a N\'eel ordering.

Figure \ref{delta_mu} shows the SP atomic 
dimerization $\delta$ at 6~K and the AF magnetic
moment $\mu_{eff}$ at 0~K as a function of Zn concentration $x$.  $\mu_{eff}$ 
is found to scale with $T_{N}$ very well, showing a linear
relation, as demonstrated in the bottom panel of Figure \ref{delta_mu}. 
As mentioned previously, the resolution-limited peak width shown in 
Fig. \ref{order_parameters}(b)
indicates a long-range AF ordering; this is true for all samples.  However,
the SP reflections exhibit line broadening for 3.4 and 4.7\% Zn-doped samples.

\section{Magnetic Excitations}
\subsection{Spin-Peierls Excitations}

In the SP state of CuGeO$_3$ excitations were observed 
to have an energy gap of 2.1 meV at $q=(0,\ 1,\ 0.5)$ [\onlinecite{nishi94}].  
Hase {\it et al.}\cite{hase96} reported that the magnetic excitations in a 
3.4\% Zn-doped CuGeO$_3$ crystal are overdamped. 
Magnetic excitations in Si-doped CuGeO$_3$ have also been recently reported 
by Regnault {\it et al.}\cite{regnault} 
where, in contrast to Hase {\it et al.}, the excitations remain well
defined (are underdamped).  Extensive measurements of 
the magnetic excitation spectra were obtained in all of the 
Zn-doped crystals of the present study.  Some preliminary results were
presented by Sasago {\it et al.}\cite{sasago}  In the subsequent examination 
of all other compositions, well-defined (underdamped) magnetic excitations 
were found.  Therefore the 3.4\% Zn-doped crystal  used by Hase 
{\it et al.}\cite{hase96} is the only sample in which overdamped excitations 
were observed.  

The measurements were extended to a wider range of $q$ within the  (0 $k$
$l$) zone to determine the dispersion along $b^*$ and $c^*$.   A summary of 
the measured dispersions is presented in Fig.\ \ref{SP_dispersion}.   
As in pure CuGeO$_3$, the dispersion is much steeper in the $c^*$ direction 
compared to $b^*$.  A small decrease in the SP energy gap magnitude 
is observed throughout the zone upon increasing Zn concentration. 
However the excitations remain well defined and are only somewhat broadened.  
This broadening is demonstrated in Fig.\ \ref{SP_excitation}, where an 
excitation profile at $q = (0,\ 1.1,\ 0.5)$ for 3.2\% Zn is compared 
with those of pure and 0.9\% Zn samples.  By convolving 
the proper resolution function  with 
a dispersion surface, the intrinsic line widths $\Gamma$ are extracted for
0, 0.9\% and 3.2\% Zn to be $\Gamma=0.22$, 0.31, and 1.09~meV, respectively.  
The solid lines in the figure are the resultant fits.  

To try to clear up the discrepancy between the present results and those
of Hase {\it et al.}\cite{hase96}, the 3.4\% Zn sample used
in that earlier study was re-examined. 
This sample was found to have a reasonably well defined peak
only at higher energy transfer.  Thus the overdamped modes observed 
in the Hase sample are limited to low energy SP magnetic excitations.  
We speculate that the AF modes may be underdamped in Hase's 3.4\% Zn sample.

\subsection{AF Excitations}

Regnault and co-workers\cite{regnault} reported finding sharp AF magnetic
excitations at energies of 0.2 meV at the magnetic zone center in their
0.7\% Si doped sample.   
Using electron spin resonance (ESR) Hase {\it et al.}\cite{haseESR} 
recently observed an antiferromagnetic resonance below $T_N=4.2$K in a
Cu$_{0.96}$Zn$_{0.04}$GeO$_3$ crystal, which indicates spin-wave 
excitations (AF modes) at the magnetic-zone center at energies of 0.11 and
0.17 meV.\cite{haseESR}  We therefore searched for these AF excitation modes
in our neutron studies of our best characterized higher doped crystal which 
has 3.4\% Zn.

In our search for AF excitations, we encountered considerable difficulties 
mostly due to various kinds of spurious peaks appearing through the
resolution function.  After experimenting with two incident neutron energies 
and three different scattering zones, the proper experimental 
window was found in which the AF excitations can be characterized in a wide
enough $q$--$E$ range.
Figure \ref{SP_AF_excitations_lowE} depicts how the AF mode is sharp 
at low $q$ values, but becomes significantly broadened at larger values 
of $b^*$.  Figure \ref{AF_excitation_a} demonstrates the dispersion and 
broadening which occurs as a function of $q$ along $a^*$.  
The AF mode along $c^*$ appears to be broad even for small $q$.  It
seems that the broadening occurs when the transfer energy, rather than
$q$, exceeds a certain value. 
It is interesting to note that the broadening at higher $q$ values  
is exactly opposite to the SP modes which are sharp at high $q$ and 
broaden dramatically at smaller $q$.  The AF excitation data are 
combined into full dispersion relations which are presented in 
Figure \ref{SP_AF_dispersion_bc}, 
along with $q$ maps showing where the measurements were carried out.

\section{Discussion}



As we have already emphasized, the AF ordering in the Cu$_{1-x}$Zn$_x$GeO$_3$
system takes place for as little as 0.42\% Zn which has a $T_N$ = 0.6~K.  
It is now established that a very small Zn 
concentration is sufficient for long range AF ordering and that this
long-range AF order is present in all Zn concentrations studied.  
These remarkable results have immediate consequences on theoretical
models.  Percolation type descriptions
are ruled out as they would predict the onset of AF only
after a certain amount of impurity islands have been formed; when the amount
of Zn is too small, long range order could not exist in such a system.
The present results instead lend weight to a theoretical model recently 
proposed by Fukuyama {\it et al.}\cite{fukuyama} which predicts a standing 
wave long
range order for arbitrarily small amounts of an impurity within a SP
system, probably even at $x=0$ for perfect crystals.  Our data of AF
ordering in a 0.42\% Zn crystal strongly support this theoretical
prediction.  

In contrast to the long-range correlations in the AF state, the SP phase 
exhibits increasing disorder with increasing Zn concentration;
the SP superlattice reflection at (1.5, 1, 1.5)
shows considerable line broadening with increasing $x$.  Two recent x-ray 
studies\cite{schoeffel,kiryukin} generally agree with our results, one
measures the line widths, and the other takes a new approach using
critical scattering.  It will be interesting to learn how the SP phase is
gradually destroyed at high $x$ values.
It may be in this aspect where the distinction between site substitution of 
Zn for Cu is different from the bond disorder in substitution 
of Si for Ge.  The ability to grow uniform, large $x$
single crystals of Cu$_{1-x}$Zn$_x$GeO$_3$ is necessary for further 
quantitative studies.  As discussed in the sample preparation description 
above, this ability is still forthcoming.

In the present study, evidence for AF and SP orderings have been presented 
along with detailed quantitative results for AF magnetic moment $\mu_{eff}$ 
and dimerization $\delta$ as a function of
Zn concentration $x$.  One important finding is that $\mu_{eff}$ scales well 
with $T_N$.  In addition, key characteristics of SP and AF
magnetic excitations have been established.  
Contrasting behaviors are found in these excitations.  The AF excitations
show a distinctive peak in the small $q$ and energy region.  As the energy
transfer increases, the AF excitation rapidly broadens, and above 1~meV it
becomes indistinguishable from the background.  On the other hand, the SP
excitations show considerable line-broadening as the Zn concentration is
increased, and this peak broadens at smaller $q$ as well.  
These results will serve as critical
tests for forthcoming theoretical models of this fascinating compound.

So far the problem has been treated as 
if the Zn content $x$ is the only parameter with which to characterize the 
Cu$_{1-x}$Zn$_x$GeO$_3$ system.  There may be other issues at play, such as
exact oxygen stoichiometry (which can have dramatic effects in other
well-known oxide systems). 
Closer inspection of Table I reveals immediately that 
$T_N$, $\mu_{eff}$, and $\delta$ are not smooth functions of $x$.  
It may, in part, be due to inaccuracies in the experiments, but
the variation is too large to be completely assigned to experimental
errors.  And the fact that the Hase 3.4\% sample
does not quite fit into the overall scheme remains a puzzle.
Another example of how much is still to be learned about the
CuGeO$_3$ system is the variance of $T_{SP}$ even for $x = 0$.  
An undoped crystal which is our most perfect crystalographically 
has a $T_{SP}$ of 13.2~K, one full degree lower than the 
accepted value of 14.2~K.

After writing this paper we became aware of a manuscript by Coad
{\it et al.}\cite{coad} which discusses elastic neutron measurements of 
CuGeO$_3$ where Cu has been replaced by Zn or Ni.  They similarly report a 
degradation of the SP state with the onset of AF ordering and their
resultant phase diagram is close, although not identical, to the one
reported here.  Overall their work and the present results lend support to each
other.

\acknowledgments
We would like to thank R.J. Birgeneau, Guillermo Castilla, and Vic
Emery for stimulating discussions.  Thanks to S. Coad, H. Fukuyama, and L.P.
Regnault for providing their manuscripts prior to publication
and  for informative private discussions.  We also are pleased to
acknowledge the expert technical help of J. Biancarosa, N. Donahue, B. Liegel,
K. Mohanty,  and R. Rothe of the BNL HFBR Physics staff.   This work was
supported in part by the U.S.- Japan Cooperative Research  Program on Neutron
Scattering, and a NEDO (New Energy and Industrial Technology Development
Organization) International Joint Research Grant.  Research at Brookhaven
National Laboratory was supported by the Division of  Materials Research at the
U.S. Department of Energy, under contract No.  DE-AC02-76CH00016.



\begin{table}
\caption{Numerical values for the SP and AF transition temperatures, and the
atomic shifts and magnetic moments for the SP and AF states, respectively,  for
all Cu$_{1-x}$Zn$_x$GeO$_3$ crystals studied.}
\begin{tabular}{c|c|d|dd|dd}
Identification&Zn&Crystal Volume&\multicolumn{2}{c}{SP}&\multicolumn{2}{c}{AF}\\
&$x$ (\%)&(cm$^3$)&$T_{SP}$ (K)&$\delta$(6K)&$T_N$ (K)&$\mu_{eff}$(0K)\\
\tableline
N 1    &0.42$\pm$0.14&0.03&13.6&0.72&0.6&0.06\\
--     &0.74         &0.01&12.5&0.60&2.4&0.14\\
N 2    &0.91$\pm$0.20&0.4 &12.5&    &2.0&0.10\\
N 6    &2.1 $\pm$0.3 &0.4 &10.2&0.26&4.2&0.24\\
N 4    &3.2 $\pm$0.4 &0.4 &10.3&0.40&4.1&0.18\\
Hase 4 &3.4          &0.04&10.0&    &4.2&0.22\\
N 8    &4.7          &0.1 &10.0&    &4.3&0.19
\end{tabular}
\label{tab}
\end{table}


\begin{figure}
\caption{Newly established magnetic phase diagram of
Cu$_{1-x}$Zn$_{x}$GeO$_{3}$ by the present neutron-scattering studies
(open circles) in combination with magnetic susceptibility measurements
(filled circles).}
\label{phase_diagram}
\end{figure}

\begin{figure}
\caption{Sketches of some of the flux-grown crystals used in the present 
study. The sections of each crystal are labeled as to what experiments 
were carried out on them:  SUS (susceptibility), EPMA (Electron Probe 
Micro-Analysis), and Neutron scattering.  Examples of the temperature 
dependence of the measured susceptibility along the c-direction is
presented in the lower part of the figure for pure, 0.91\% and 3.2\% Zn-doped
crystals.}
\label{samples}
\end{figure}

\begin{figure}
\caption{(a) Intensities of SP and AF superlattice peaks as a function of
temperature for a 3.2\% Zn-doped crystal.  The two transition temperatures are
clearly observable and are indicated on in the plot. (b) 
Temperature dependence of an AF superlattice peak for a 0.42\% Zn-doped 
crystal.  The inset shows a peak profile of the AF superlattice peak,
indicating that the width is resolution limited.}
\label{order_parameters}
\end{figure}


\begin{figure}
\caption{(a) Dimerization ($\delta$) at 6~K as a function of Zn concentration
($x$).  (b) AF magnetic moment ($\mu_{eff}$) extrapolated to 0~K as a
function of $x$.  (c)  $\mu_{eff}$ correlates linearly with $T_N$.}
\label{delta_mu}
\end{figure}

\begin{figure}
\caption{The measured dispersions of the SP magnetic excitation in the $b^*$
and $c^*$ directions.  The different symbols represent different Zn
concentrations as denoted in the legend.}
\label{SP_dispersion}
\end{figure}

\begin{figure}
\caption{SP excitations of Cu$_{1-x}$Zn$_{x}$GeO$_{3}$ for $x=0,\,
0.009,\, 0.032$ above $T_{N}$ of each sample.  Data are fitted with a
scattering function described in the text with a proper resolution correction,
in order to extract an intrinsic energy width $\Gamma$ of the excitation. }
\label{SP_excitation}
\end{figure}


\begin{figure}
\caption{SP as well as AF excitations for a 3.2\% Zn-doped sample measured at
$E_{i}=5.0$~meV.  The AF mode can be seen clearly, its energy width is
resolution limited at small $q$, but becomes significantly broadened at higher
$q$.  Lines are guides to the eye.}
\label{SP_AF_excitations_lowE}
\end{figure}

\begin{figure}
\caption{Energy spectra of AF excitations for a 3.2\% Zn-doped sample measured
at $q$ positions along the $a^*$ direction, which has considerably less 
dispersion than in the $b^*$ or $c^*$ directions.}
\label{AF_excitation_a}
\end{figure}


\begin{figure}
\caption{The measured dispersions of the SP as well as AF magnetic excitations
in the $b^*$, $c^*$ and $a^*$ directions.  Insets shows where in $q$-space the 
scans were obtained.}
\label{SP_AF_dispersion_bc}
\end{figure}



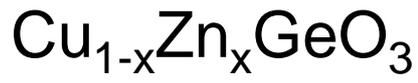
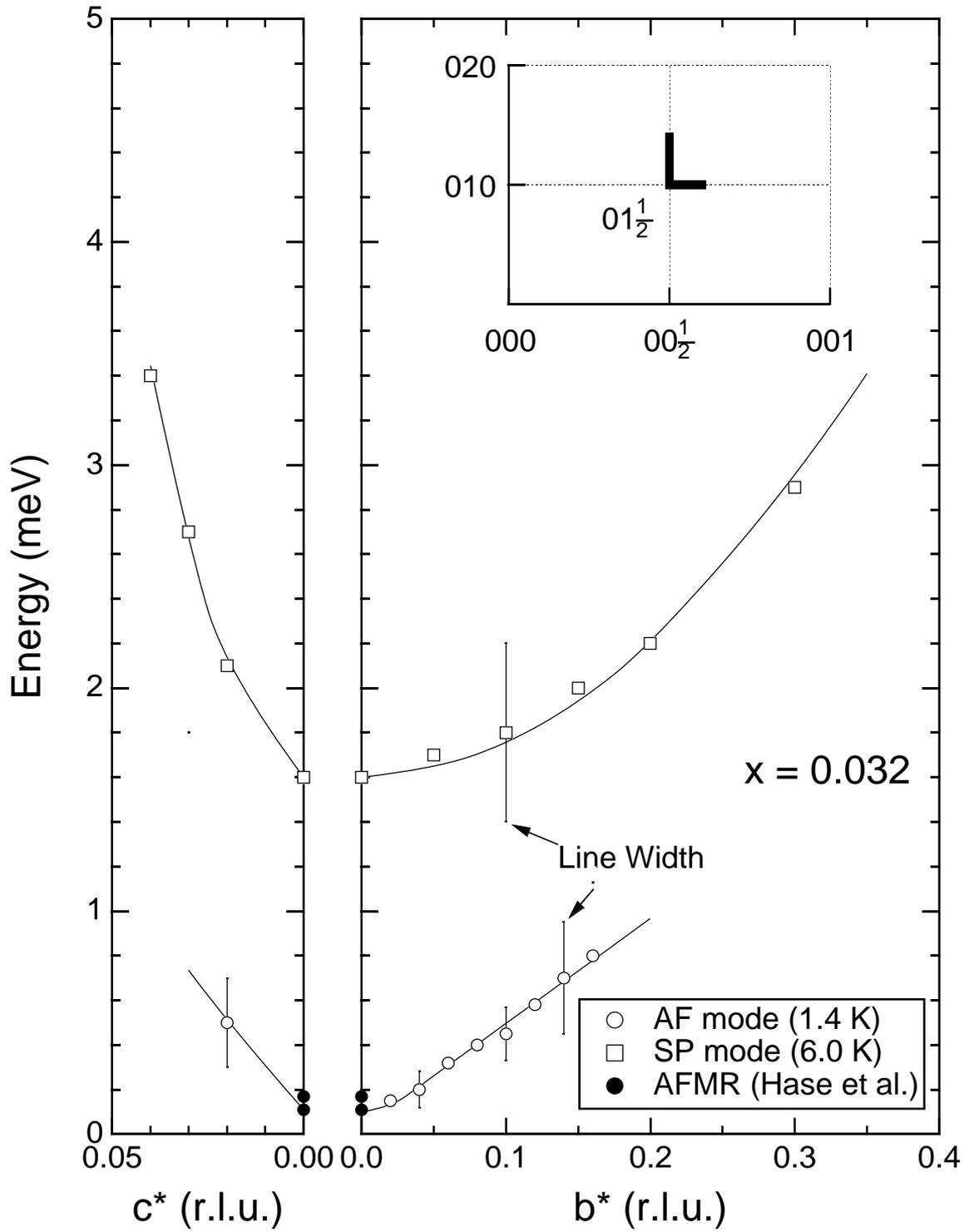
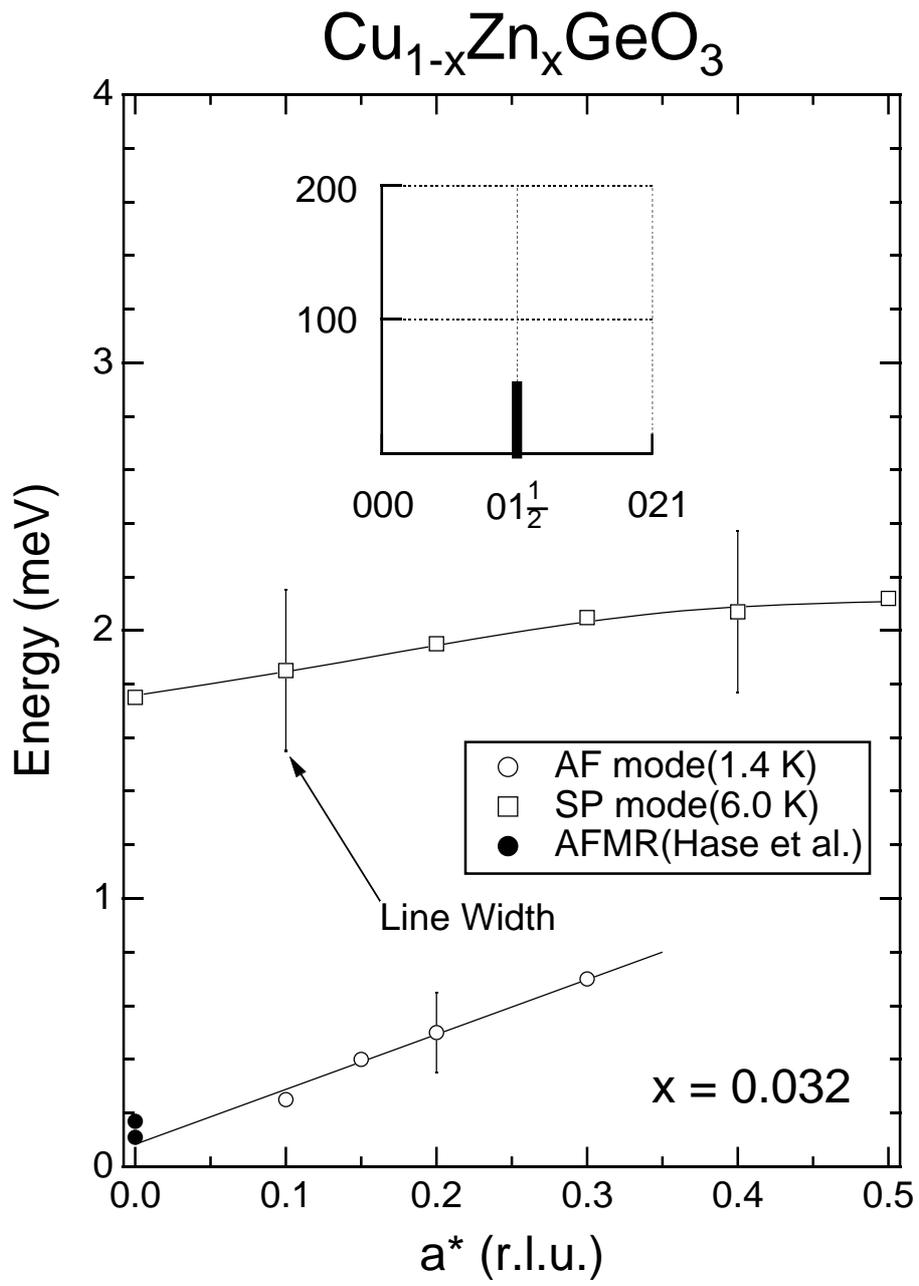
\end{document}